\documentclass[fleqn,11pt]{wlscirep}
\usepackage[utf8]{inputenc}
\usepackage[T1]{fontenc}
\usepackage[utopia]{mathdesign}
\usepackage[OMLmathrm,OMLmathbf]{isomath}

\newcommand{\beginsupplement}{%
        \setcounter{table}{0}
        \renewcommand{\thetable}{S\arabic{table}}%
        \setcounter{figure}{0}
        \renewcommand{\thefigure}{S\arabic{figure}}%
     }

\usepackage[numbers]{natbib}
\usepackage{setspace}
\def\LAOSTO{LaAlO$_3$/SrTiO$_3$ }
\def\STO{SrTiO$_3$ }
\def\LAO{LaAlO$_3$ }

\title{Engineered spin-orbit interactions in LaAlO$_3$/SrTiO$_3$-based 1D serpentine electron waveguides}

\author[1,2]{Megan Briggeman}
\author[1,2]{Jianan Li}
\author[3]{Mengchen Huang}
\author[4]{Hyungwoo Lee}
\author[4]{Jung-Woo Lee}
\author[4]{Kitae Eom}
\author[4]{Chang-Beom Eom}
\author[1,2]{Patrick Irvin}
\author[1,2,*]{Jeremy Levy}
\affil[1]{University of Pittsburgh, Department of Physics and Astronomy, Pittsburgh, PA 15260, USA}
\affil[2]{Pittsburgh Quantum Institute, Pittsburgh, PA 15260, USA}
\affil[3]{University of California-Santa Barbara, Department of Physics, Santa Barbara, CA 93106, USA}
\affil[4]{University of Wisconsin-Madison, Department of Materials Science and Engineering, Madison, WI 53706, USA}

\affil[*]{corresponding author jlevy@pitt.edu}

\begin{abstract}
The quest to understand, design, and synthesize new forms of quantum matter guides much of contemporary research in condensed matter physics. One-dimensional (1D) electronic systems form the basis for some of the most interesting and exotic phases of quantum matter \citep{Giamarchi2003}. 
The variety of experimentally-accessible ballistic 1D electronic systems is highly restricted, and furthermore these systems typically have few tuning parameters other than electric and magnetic fields.
However, electron waveguides \citep{Annadi2018} formed from two-dimensional (2D) LaAlO$_3$/SrTiO$_3$ heterointerfaces exhibit remarkable 1D properties, including ballistic multi-mode transport and strong attractive electron-electron interaction \citep{Cheng2015,Cheng2016}, but these systems conspicuously lack strong or tunable spin-orbit interactions.
Here we describe a new class of quasi-1D nanostructures, based on \LAOSTO electron waveguides, in which a sinusoidal transverse spatial modulation is imposed.  
Nanowires created with this ``serpentine'' modulation display unique dispersive features in the subband spectra,
namely (1) a significant shift ($\sim$ 7 tesla) in the spin-dependent subband minima, and (2) fractional conductance plateaus, some of which are continuously tunable with a magnetic field.  
The first property can be understood as an engineered spin-orbit interaction associated with the periodic acceleration of electrons as they undulate through the nanowire (ballistically), while the second property signifies the presence of enhanced electron-electron scattering in this system due to the imposed periodic structure.  
The ability to engineer these interactions in quantum wires contributes to the tool set of a 1D solid-state quantum simulation platform.
\end{abstract}
\begin{document}

\flushbottom
\maketitle
\thispagestyle{empty}
\section*{Introduction}

\indent 

One approach to the grand challenge of understanding new states of quantum matter is through ``quantum simulation'', the creation of a highly configurable many-body quantum system is developed in which its Hamiltonian description can be related to relevant physical models \cite{Feynman1982,Cirac2012,Georgescu2014}.
Quantum simulation necessarily requires a physical platform that can be configured to match or approximate the system of interest.
Among the many quantum systems being developed for this purpose, ultracold atoms trapped within standing waves of light \citep{Jaksch2005} have been particularly successful, in large part because the model Hamiltonians are well characterized and based upon a fundamental understanding of the constituent atomic systems. For example, hyperfine states of trapped ions have been greatly successful in simulating classes of spin chains \citep{Bohnet2016,Zhang2017}.  Superconducting networks can also be used to simulate a wide range of Hamiltonians \cite{Houck2012}, while atom-scale manipulation (e.g., donor atoms in silicon \citep{Salfi2016} or arrangements of CO molecules \citep{Gomes2012}) has successfully emulated band structure and topological phases.

The type of quantum systems that can be explored in a quantum simulator is often limited by the available interactions of the host material. To increase the available physical interactions, a variety of pseudo-magnetic fields \citep{Abo-Shaeer2001,Engels2003}, gauge fields \citep{Lin2009}, and spin-orbit interactions \citep{Lin2011}, can be added.  Inter-particle interactions can be controlled in a variety of ways, e.g., via Feshbach resonance in atomic systems, or by coupling to a polarizable medium \citep{Hamo2016}.

The complex-oxide \STO possesses a wide range of gate-tunable properties that include superconductivity, magnetism, ferroelectricity, and ferroelasticity \citep{Pai2018a}.  Using a conductive atomic-force microscope (c-AFM) lithography technique, the \LAOSTO interfacial conductivity (and related properties) can be programmed with a precision of two nanometers \citep{Cen2008,Cen2009}, comparable to the mean separation between electrons.  The combination of a rich palette of intrinsic properties and the ability to form complex nanostructures provides a suitable foundation for the creation of a 1D quantum simulation platform.

\begin{figure}
    \centering
    \includegraphics[width=\columnwidth]{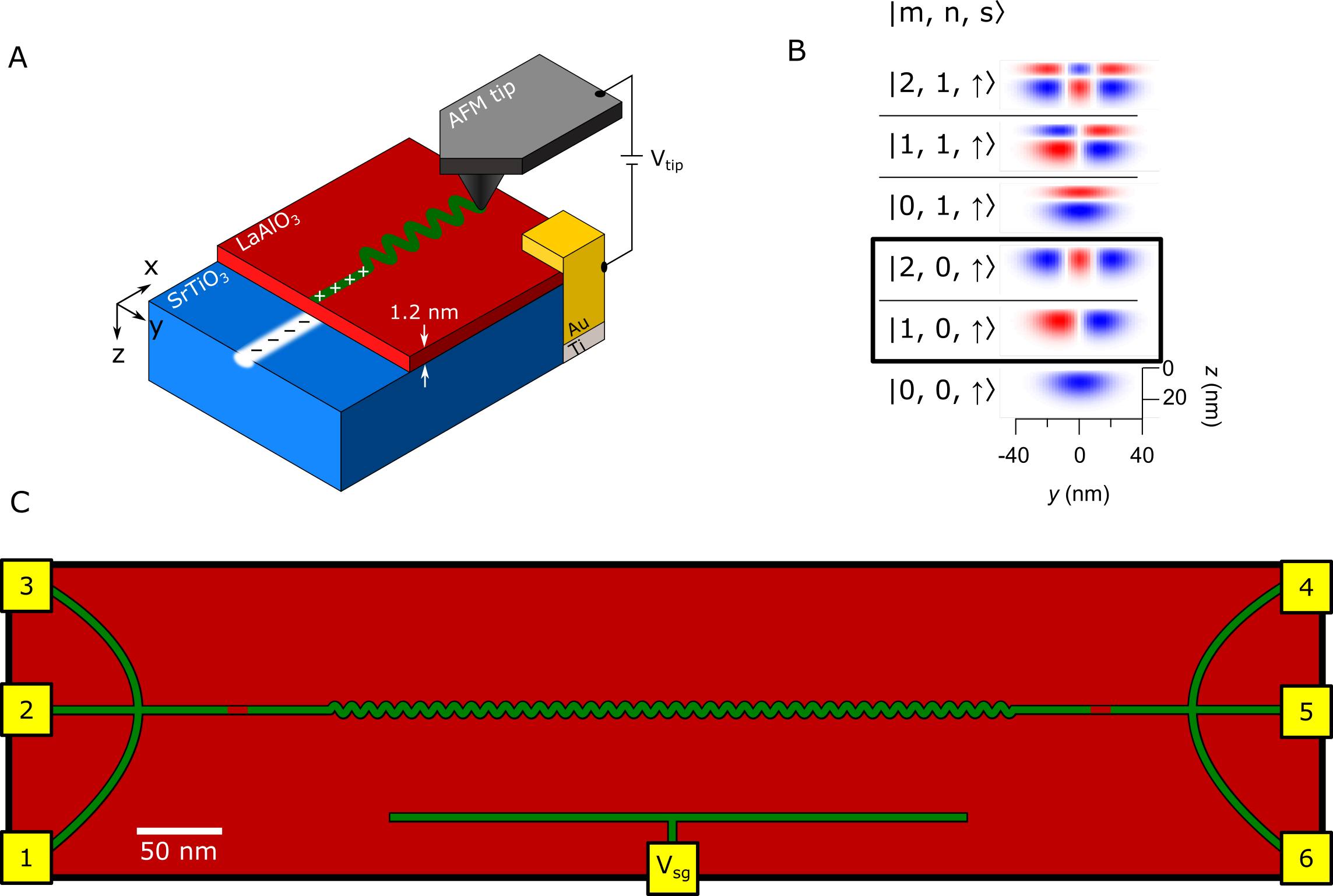}
    \caption{\textbf{Conductive AFM writing and device schematic.}  \textbf{(A)} Nanowires created at the \LAOSTO interface using c-AFM lithography.  A positive bias applied to the AFM tip protonates the surface, causing electrons to accumulate at the interface.  1D serpentine superlattice devices are created by laterally modulating the tip position on the \LAO surface. \textbf{(B)} Representative wavefunctions calculated for an electron waveguide device with vertical, lateral, and spin degrees of freedom \citep{Annadi2018}.  The serpentine motion of the superlattice couples the ground state of the waveguide with different lateral modes of the waveguide (modes circled in black).  \textbf{(C)} Schematic for the 1D serpentine superlattice devices.  C-AFM written paths (green lines) represent the device and are connected to interface electrodes (yellow).  The serpentine lateral modulation is bracketed by highly transparent tunnel barriers similar to those used to create electron waveguide devices \citep{Annadi2018}.  The voltage/current leads are used to take a 4-terminal measurement of the device.  A local side gate is also created using c-AFM lithography.  A voltage applied to the gate ($V_\mathrm{sg}$) changes the chemical potential of the the device.  }
    \label{fig:fig1}
\end{figure}

A useful starting point for developing programmable 1D quantum systems is the \LAOSTO electron waveguide \citep{Annadi2018}.  These devices exhibit highly-quantized ballistic transport, in which the conductance is quantized in units of $e^2/h$, where $e$ is the electron charge and $h$ is the Planck constant.  Each of the $N$ occupied 1D subbands (arising due to vertical, lateral, and spin degrees of freedom) contributes one quantum of conductance to the total conductance $G=Ne^2/h$.  A variety of correlated electronic phases have been identified, including a paired liquid phase \cite{Cheng2015,Annadi2018}, re-entrant pairing \cite{Annadi2018}, and a family of emergent composite electron liquids comprised of bound states formed from 2,3,4,... electrons \citep{Briggeman2019}.  The calculated wavefunctions of a representative electron waveguide device are shown in Figure \ref{fig:fig1}C, where the state $|m,n,s\rangle$ is indentified by its quantum numbers $m$, $n$, and $s$ that characterize the transverse orbital and spin degrees of freedom.  Much of the unusual transport characteristics come from interactions between these various electronic subbands.  

One property that appears to be lacking (or weak) in \LAOSTO electron waveguides is spin-orbit coupling.  Gate-tunable spin-orbit coupling has been reported at the 2D \LAOSTO interface \citep{Shalom2010,Caviglia2010}; however, detailed modeling of the subband spectra have ruled out such interactions for the most part in 1D quantum wires \citep{Annadi2018}.  Strong spin-orbit interactions are believed to be the ``missing ingredient'' in efforts to create Majorana zero modes \cite{Lutchyn2010,Oreg2010} in these 1D quantum wires.  A reasonable goal is therefore to engineer spin-orbit interactions in quantum wires, using the nanoscale control enabled by c-AFM lithography.

Here we present transport experiments on ballistic electron waveguides that are perturbed by a periodic transverse (``serpentine") spatial modulation (Figure \ref{fig:fig1}).  Conductive nanostructures are created at the \LAOSTO interface using a positively-biased c-AFM tip placed in contact with the \LAO surface, locally switching the interface to a conducting state because of local protonation of the \LAO surface \citep{Bi2010,Brown2016}. We perturb the electron waveguide structure by superimposing a periodic transverse modulation to the device (Figure \ref{fig:fig1}).  The path for a sinusoidal waveguide oriented along the $x$ direction is given by $y(x)=y_0 +y_k \sin(k x)$, where $y_0$, $y_k$ and $k$ are parameters that can be programmed.  The impact of this modulation on the transverse mode, expressed using the basis of unperturbed states ($|m,n,s\rangle$), is expected to be dominated by the $|1,0,s\rangle$ state, with a higher correction from the $|2,0,s\rangle$ state (Figure \ref{fig:fig1}B).  

\section*{Results}

\begin{figure}
    \centering
    \includegraphics[width=\columnwidth]{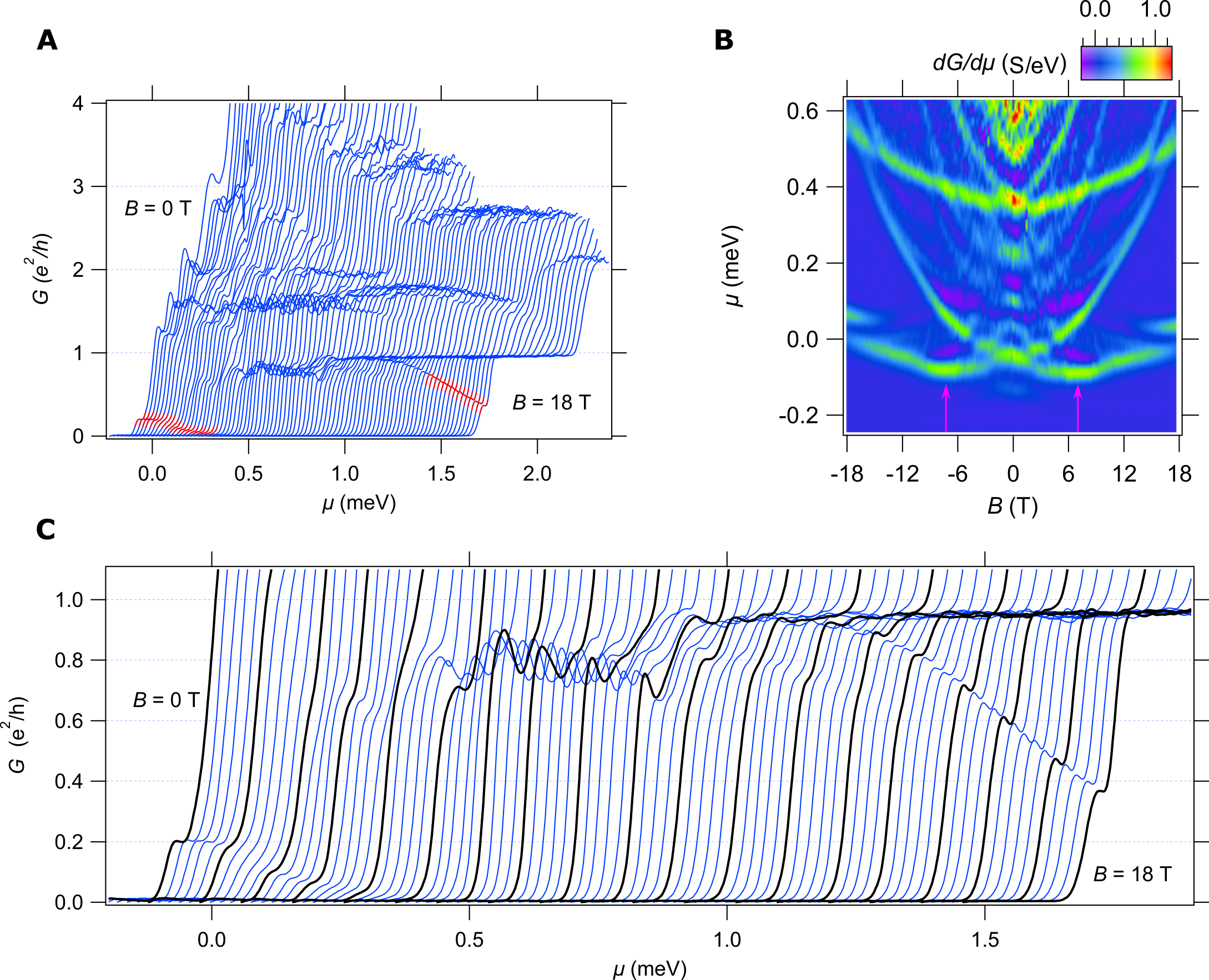}
    \caption{\textbf{Magnetotransport for serpentine superlattice Device A.}  \textbf{(A)} Conductance, $G$, plotted as a function of chemical potential $\mu$ and applied magnetic field $B$ from 0 T (leftmost) to 18 T (rightmost). Curves are offset for clarity. Fractional conductance features below the $1~e^2/h$ plateau are hightlighted in red. \textbf{(B)} Transconductance $dG/d\mu$ as a function of magnetic field $B$ and chemical potential $\mu$.  Light (red/yellow/green) regions indicate increasing conductance, i.e. when new subbands become available.  Dark blue regions indicate zero transconductance or conductance plateaus.  Purple regions are regions of negative transconductance and indicate decreasing conductance.  The minima of the lowset subband occurs at finite $B$ field values, highlighted with pink arrows. \textbf{(C)} Zoom in of the fractional conductance features below the $1~e^2/h$ plateau. Curves at 1 T intervals are highlighted in black.}
    \label{fig:fig2}
\end{figure}

Four-terminal magnetotransport data for a serpentine superlattice (Device A) is shown in Figure \ref{fig:fig2}.  Measurements are taken at or near the base temperature of a dilution refrigerator ($T=25$ mK), as a function of out-of-plane magnetic field $|B|\leq 18$ T and chemical potential $\mu$, which is controlled by the voltage on a local side gate.  Device design parameters are summarized in Table \ref{tab:devices}.  The four-terminal conductance $G$ as a function of $\mu$ and $B$ (Figure \ref{fig:fig2}A) shows quantized plateaus that result from Landauer quantization, similar to what is observed for unperturbed (straight) electron waveguides.  In addition, the device shows a number of fractional conductance plateaus. Two features are highlighted in red, and shown in expanded detail in Figure \ref{fig:fig2}C. The conductance value of these fractional feature evolves down from the $\sim 1~e^2/h$ plateau, reaching a value of $\sim 0.4~e^2/h$ at $B=18$ T.  A smaller fractional conductance feature, $\sim 0.2~e^2/h$ near zero magnetic field, remains stable until about $B=1$ T, and then decreases in magnitude with increasing $B$ field before disappearing at $B\approx5$ T.  Several fractional conductance states are observable at higher overall conductances, which are also  tunable with a magnetic field, e.g., a feature between 1.5 $e^2/h$ and 1.8 $e^2/h$.  In many instances, there is significant overshoot (resulting in parameter regimes for which $dG/d\mu<0$) before a plateau is reached.

Transconductance maps $dG/d\mu$, when plotted versus $B$ and $\mu$ (Figure \ref{fig:fig2}B) provide additional insight into the transport characteristics of these serpentine superlattices.  
In the color scheme, bright green/yellow/red regions ($dG/d\mu>0$) represent increases in conductance that generally correspond to introduction of new 1D subbands.  Dark blue regions ($dG/d\mu\approx 0$) represent flat conductance plateaus, while purple regions ($dG/d\mu<0$) correspond to regions of negative transconductance.  
One standout feature of the transconductance is a shifting of the lowest subband minima to a non-zero value of the magnetic field ($B=-7.4$ T and $B=7.1$ T).  In this range of magnetic fields, a large overshoot in the conductance is also observed, followed by a region of decreasing conductance.  A second feature, observed in two ranges of magnetic field, is the existence of magnetic-field-tunable plateaus, seen near zero magnetic field and in the range $12-18$ T.

Qualitatively similar behavior is also observed for Device B (Figure \ref{fig:deviceB}), which is created in a similar manner.  Unlike Device A, Device B shows excess conductance characteristic of superconductivity near zero magnetic field rather than a fractional conductance plateau (Figure \ref{fig:0T_IV}). The transconductance map (Figure \ref{fig:deviceB}B) also shows some asymmetries in field which are due to slow temporal drifting of the chemical potential.

\begin{figure}
    \centering
    \includegraphics[width=\columnwidth]{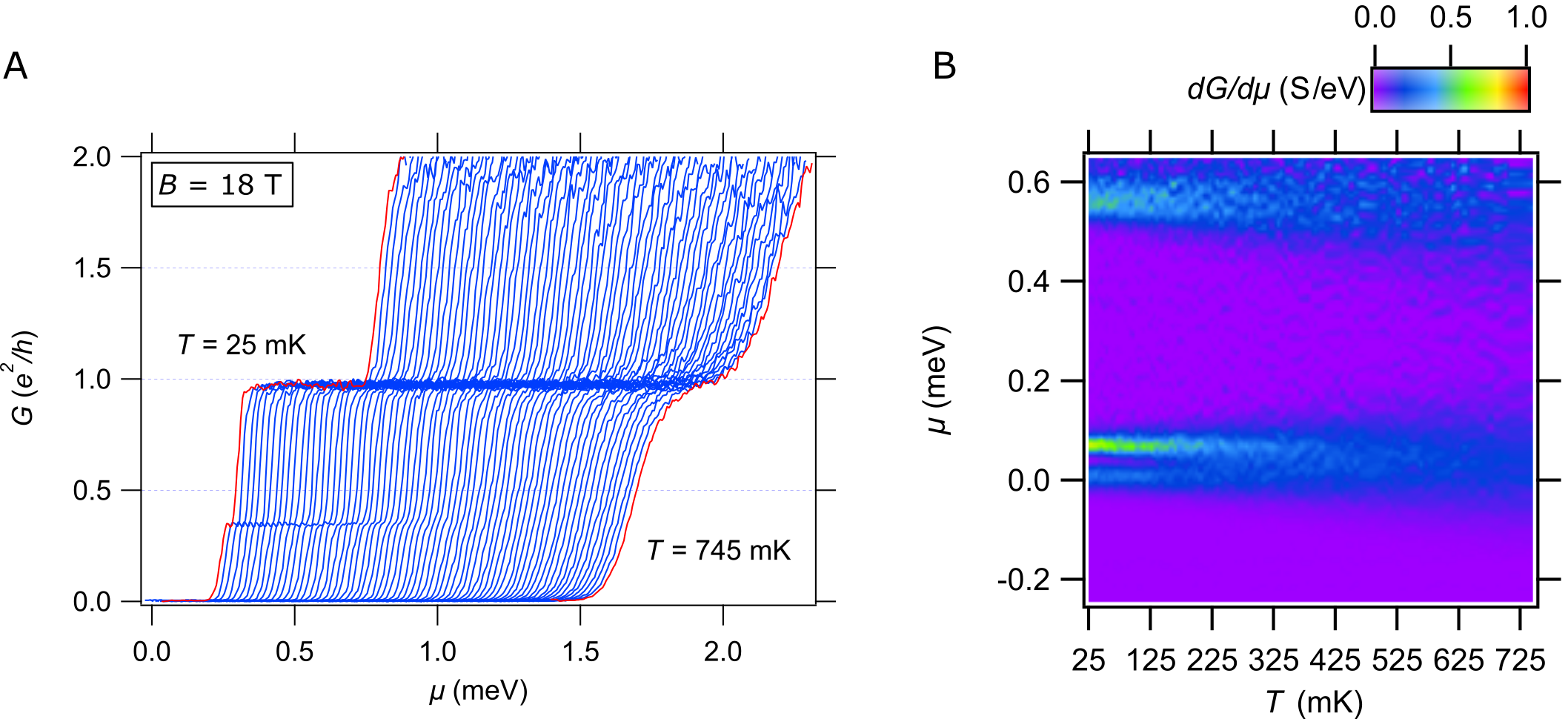}
    \caption{\textbf{Temperature dependence of Device A.} \textbf{(A)} Conductance $G$ as a function of chemical potential $\mu$ at $B=18~\mathrm{T}$ for temperatures from $T=745$ mK to 25 mK.  Temperatures are measured at the mixing chamber of the dilution refrigerator.  \textbf{(B)} Transconductance $dG/d\mu$ vs temperature $T$ and chemical potential $\mu$.  The fractional conductance feature disappears at around 200 mK, while the $1~e^2/h$ conductance plateau is still visible at 745 mK, the highest measured temperature.}
    \label{fig:fig3}
\end{figure}

The high magnetic field $B=18~\mathrm{T}$ fractional conductance feature in Device A is examined as a function of temperature $T$ and $\mu$ (Figure \ref{fig:fig3}). At the lowest temperature ($T=25~\mathrm{mK}$), the fractional feature appears as a dip in the conductance at around $0.4~e^2/h$. The dip flattens out with increasing temperature until it disappears at  $\sim200~\mathrm{mK}$. The $1~e^2/h$ plateau however persists up to 750 mK, the highest temperature that was measured.  Figure \ref{fig:fig3}B shows the transconductance map $dG/d\mu$ as a function of $\mu$ and $T$. All temperatures reported were measured at the mixing chamber stage of the dilution refrigerator.

\begin{figure}
    \centering
    \includegraphics[width=\columnwidth]{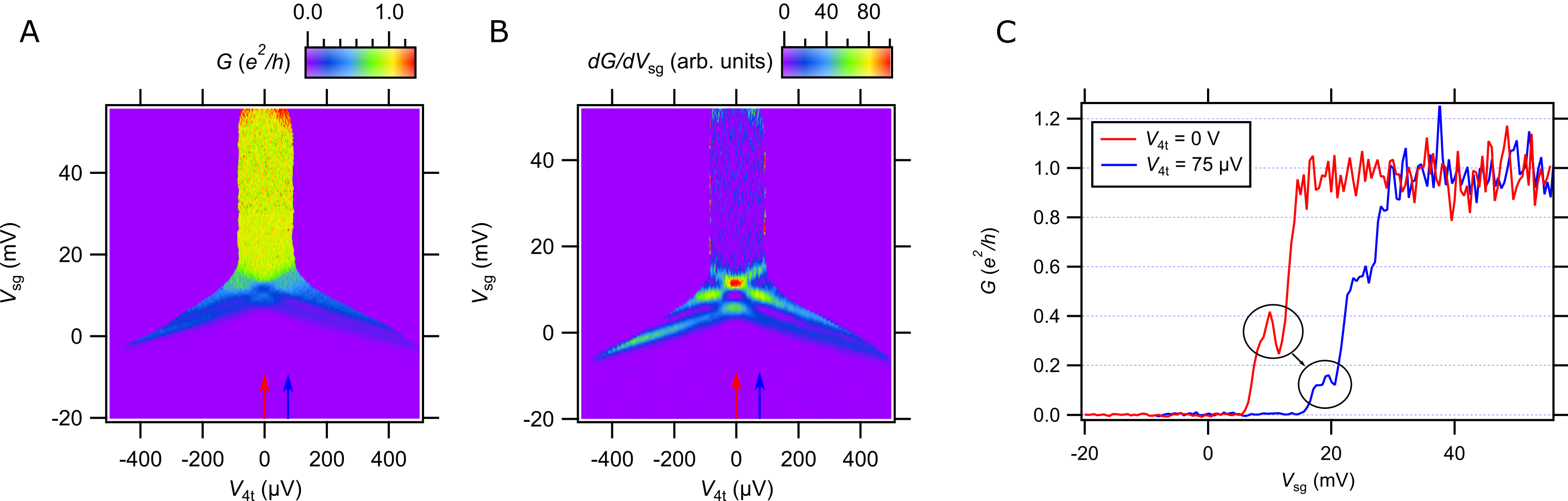}
    \caption{\textbf{Finite-bias spectroscopy for Device A.}  \textbf{(A)} Conductance map for Device A at $B=-18~\mathrm{T}$ as a function of 4-terminal voltage $V_{\mathrm{4t}}$ and side gate voltage $V_{\mathrm{sg}}$.  Conductance linecuts at $V_{\mathrm{4t}}=0~\mathrm{V}$ and 75 $\mathrm{\mu V}$ are shown in \textbf{(C)}.  \textbf{(B)} Transconductance map corresponding to panel (A).  Red and blue arrows indicate locations of line cuts in (C). The transconductance maps shows the diamond feature characteristic of ballistic transport.  Conductance linecuts show fractional conductance features below the $1~e^2/h$ plateau.  At zero-bias the conductance feature appears at $\sim0.3~e^2/h$ and at finite-bias at around half that value $\sim0.15~e^2/h$. Curves are offset for clarity. All data taken at $T=25$ mK.}
    \label{fig:fig4}
\end{figure}

\begin{figure}
    \centering
    \includegraphics[width=\columnwidth]{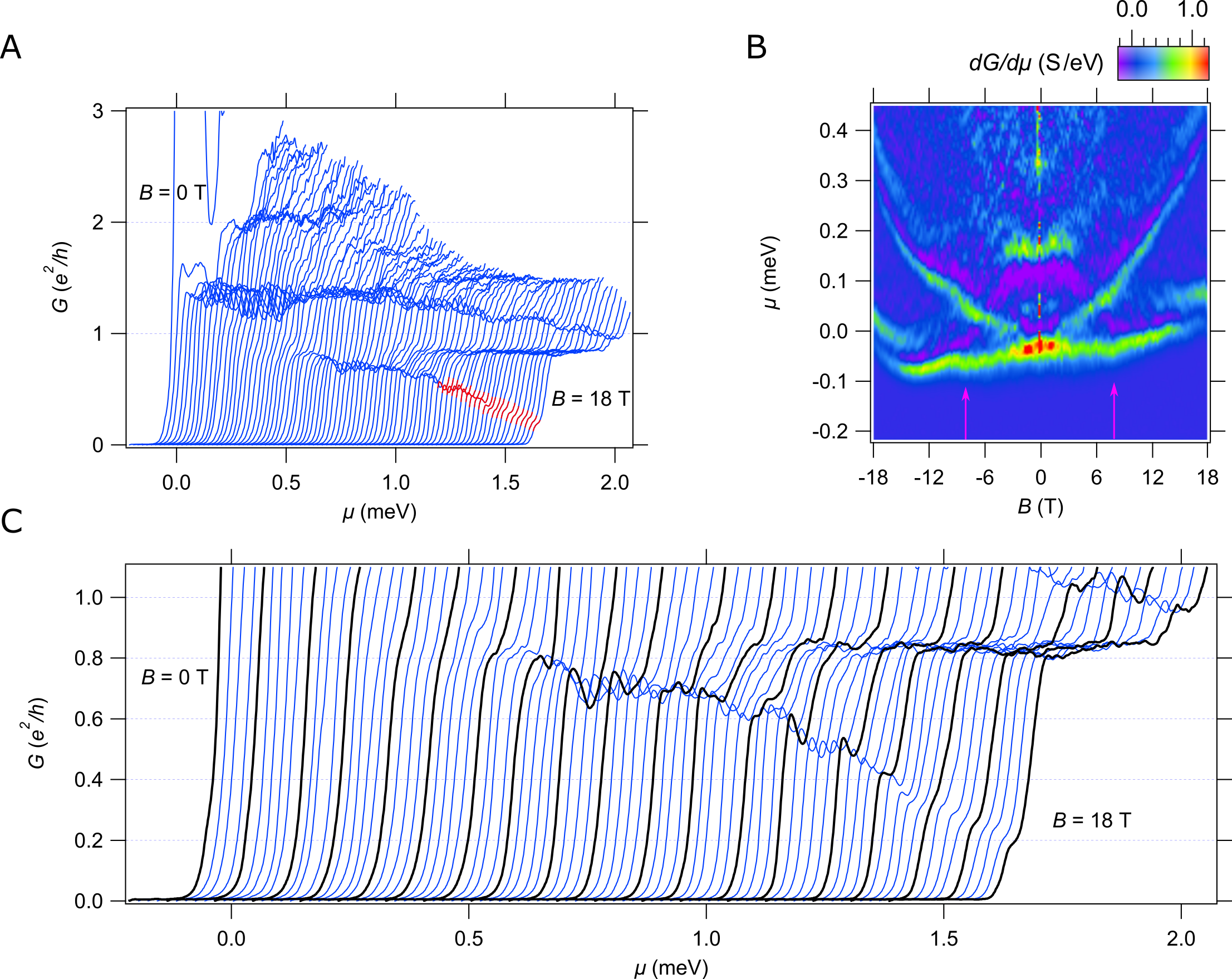}
    \caption{\textbf{Magnetotransport for serpentine superlattice Device B.} \textbf{(A)}. Conductance $G$ as a function of chemical potential $\mu$ for device B.  Curves are at different applied out-of-plane magnetic field values from 0 T to 18 T.  A fractional conductance feature is observed below the $1~e^2/h$ plateau at high magnetic fields (highlighted in red). \textbf{(B)} Transconductance map as a function of magnetic field $B$ and chemcial potential $\mu$. An overall drift while gating effects the subband structure of this device.  \textbf{(C)} Zoom in of the high field fractional conductance feature. All data taken at $T=25$ mK.}
    \label{fig:deviceB}
\end{figure}

\begin{figure}
    \centering
    \includegraphics[width=\columnwidth]{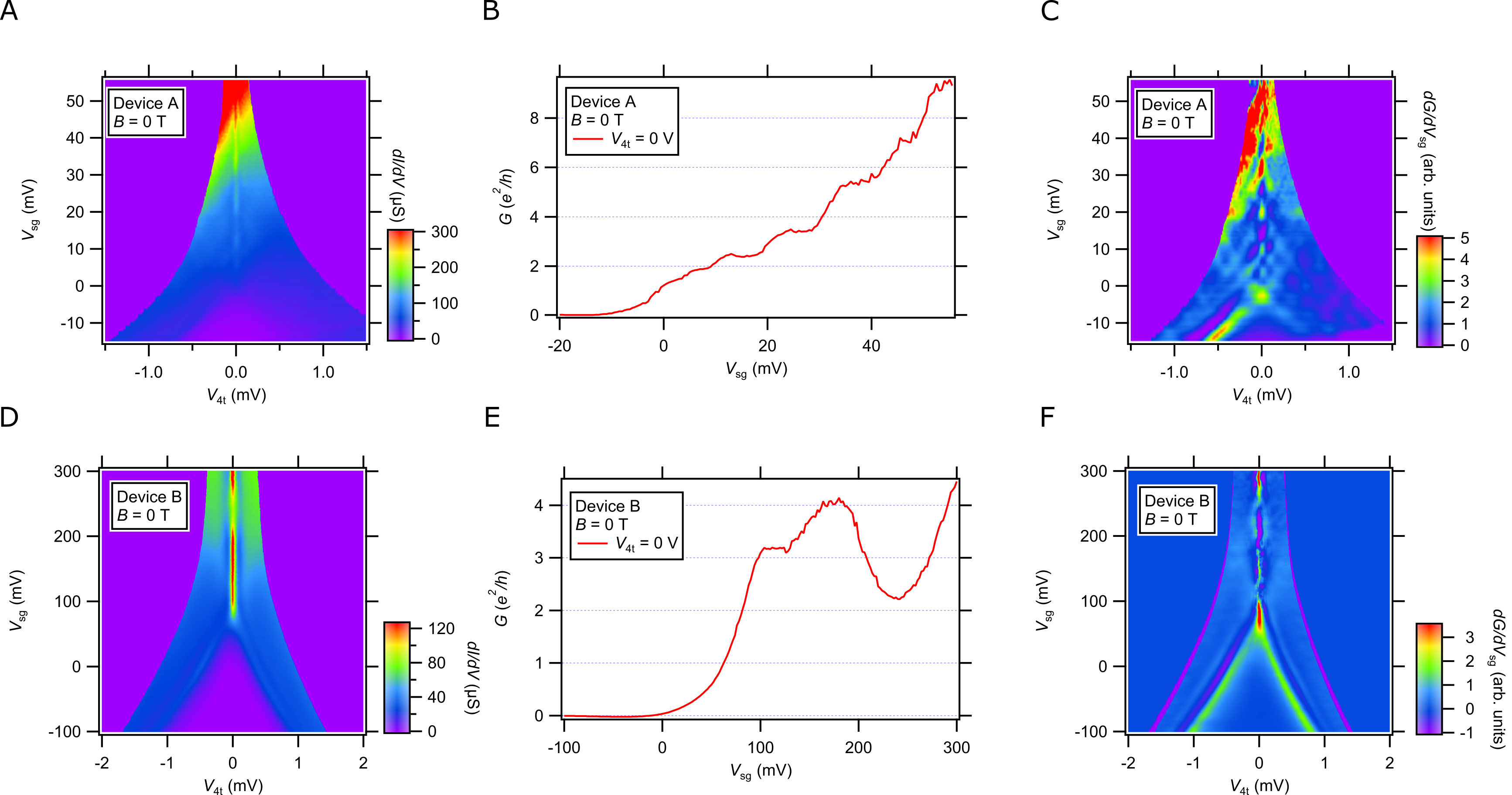}
    \caption{\textbf{Finite-bias spectroscopy for Devices A and B}  \textbf{(A)} Conductance $dI/dV$ as a function of four-terminal voltage $V_{\mathrm{4t}}$ and side gate voltage $V_{\mathrm{sg}}$ for Device A at $B=0$ T. \textbf{(B)} Zero bias ($V_{\mathrm{4t}}=0$ V) conductance line cut of Device A. \textbf{(C)} Transconductance map $dG/dV_{\mathrm{sg}}$ as a function of four-terminal voltage and side gate voltage. \textbf{(D)} Conductance $dI/dV$ for Device B.  This device shows signatures of superconductivity near zero bias. \textbf{(E)} Conductance linecut for Device B. \textbf{(F)} Transconductance $dG/dV_{\mathrm{sg}}$ for Device B. All data taken at $B=0$ T.}
    \label{fig:0T_IV}
\end{figure}

The conductance and transconductance maps as a function of four-terminal voltage $V_{\mathrm{4t}}$ and side gate voltage $V_{\mathrm{sg}}$ are shown in Figures \ref{fig:fig4}A and C, respectively.  Linecuts of the conductance at zero bias and at a finite bias of $V_{\mathrm{4t}}=75~\mathrm{\mu V}$ are shown in Figures \ref{fig:fig4}B and D, respectively.  The transconductance map shows a diamond structure characteristic of transport for ballistic systems \citep{Glazman1989,Patel1990}.  Large finite biases give rise to half-plateaus due to unevenly populated subbands.  The diamond structure implies that the conductance features are not the result of energy-dependent transmission through the device.  The value of the feature at zero bias is around $0.3~e^2/h$ and is reduced to approximately half that value at finite bias, $0.15~e^2/h$.

\section*{Discussion}

By perturbing the path of a ballistic electron waveguide, we find that it is possible to modify the spin-dependent subband structure in a manner that is consistent with an engineered spin-orbit interaction and results in the creation of new fractional conductance states.
The origin of the spin-orbit interactions can be understood in a few different ways.  The most naive explanation recognizes that the serpentine path of the electrons exposes propagating electrons, with momentum $\vec{k}=k \hat{x}$, to a spatially periodic alternating electric field, $\vec{E}_{\mathrm{eff}}(x)= E_0 \mathrm{sin} (k x) \hat{y}$, which, in the moving reference frame of the electrons, corresponds to an alternating effective magnetic field $\vec{B}_{\mathrm{SO}}\propto \vec{k}\times \vec{E}_{\mathrm{eff}}$ that is aligned with the $\hat{z}$ axis.  The resulting spin-orbit field is expected to cause a spin-dependent energy shift of the subband minima by $\pm |\vec{B}_{\mathrm{SO}}|$, consistent with our experimental findings (Figure \ref{fig:fig2}A).  A more sophisticated approach would take into account the fact that, in the basis of the unperturbed (straight) nanowire, the matrix elements that lead to hybridization with other lateral modes (highlighted in Figure \ref{fig:fig2}B) are significantly enhanced in the serpentine waveguide, and hence the magnitude of the Rashba spin-orbit interaction is correspondingly enhanced.  Details of such a calculation are beyond the scope of the current manuscript, but are nevertheless important for describing numerically-accurate models of the engineered spin-orbit interactions.

The second main experimental observation concerns the fractional conductance plateaus which exist both in zero magnetic field as well as higher magnetic fields.  In some cases the plateaus are preceded by conductance peaks.  
Fractional conductance states are typically an indication of strong electron-electron interactions.  Well known examples are the fractional quantum hall effect \citep{Tsui1982}, and the ``0.7'' anomaly which is commonly observed in quantum point contact devices \citep{Thomas1996} and which is attributed to strong interactions \citep{Pepper2008}.  There have been theoretical predictions of fractional conductance states in clean 1D systems with arbitrarily many channels and strong (repulsive) electron-electron interactions \citep{Oreg2014,Shavit2019}.  Oreg \textit{et al.} \citep{Oreg2014} studied 1D wires with spin-orbit coupling and found that in wires with strong interactions and low densities, fractional quantized conductances were predicted.  This theory predicts a plateau at $0.2e^2/h$ which usually requires broken time reversal symmetry which is inconsistent with the observed feature at 0.2 $e^2/h$ in our system at $B=0$ T.
These fractional states arise due to correlated scattering processes from different channels that can lead to fractional conductance plateaus at various rational fractions.  The underlying scattering process relies on two ingredients: (1) multiple channels from which to scatter in the forward and reverse directions, and (2) strong electron-electron interactions that support these correlated exchange of momenta.  
The Shavit-Oreg theory \citep{Shavit2019} was recently compared to experiments from Kumar \textit{et al.} \cite{Kumar2019}, in which fractional conductance plateaus were observed in 1D GaAs-based quantum wires.  While both the GaAs-based 1D wire and the \LAOSTO nanowires are ballistic, the nature of the electron-electron interactions is fundamentally different for these two materials.  That is to say, in GaAs it is repulsive, while in \LAOSTO it is strongly attractive \citep{Cheng2015,Cheng2016,Annadi2018,Briggeman2019}.  Many of the devices show signs of superconductivity at $B=0$ T (Figure \ref{fig:0T_IV}D), indicating that the interactions in these devices are indeed attractive.  The fact that such similar phenomena are identified in both systems is interesting and raises the question: to what extent can the \LAOSTO system be modeled as a system with effectively repulsive interactions?  It is known theoretically that there is a mapping between the repulsive-U and attractive-U Hubbard models \citep{Li2017}.  Perhaps this mapping can be used to understand the attractive side of the phase diagram.

Quasi-1D superlattice devices with engineered properties may provide a building block for more complex quantum systems, for example, topological phases in coupled arrays of quantum wires \citep{Kane2002,Klinovaja2016,Kane2017}.  It may also be possible to observe Majorana fermions in this system \cite{Mourik2012}.  

With the engineering of a spin-orbit interaction we may have the missing ingredient in \LAOSTO nanowire devices.  It is also worth emphasizing that these are real electronic materials and not just simulations, with engineerable properties that can be integrated with other materials or incorporated into real electronic devices.  

\section*{Methods}

\LAOSTO samples were grown using pulsed laser deposition (PLD) described in more detail elsewhere \citep{Cheng2011}.  Electrical contact was made to the interface by ion milling and depositing Ti/Au electrodes.  C-AFM writing was performed by applying a voltage bias between the AFM tip and the interface, with a 1 G$\mathrm{\Omega}$ resistor in series.  Writing was performed in 30-40\% relative humidity using a Asylum MFP-3D AFM.  Written samples were then transferred into a dilution refrigerator.  Four-terminal measurements were performed using standard lock-in techniques at a reference frequency of 13 Hz and an applied AC voltage of 100 $\mathrm{\mu}$V.

\bibliographystyle{apsrev4-1}


%

\section*{Acknowledgements}

JL acknowledges a Vannevar Bush Faculty Fellowship, funded by ONR (N00014-15-1-2847). Work at the University of Wisconsin was supported by funding from the DOE Office of Basic Energy Sciences under award number DE-FG02-06ER46327 (C.B.E).

\section*{Author contributions statement}

M.B., P.I., and J.Levy conducted the experiments.  J.Li and M.H. processed the samples. H.L., J.-W.L., K.E. and C.-B. E. synthesised the thin films and performed structural and electrical characterizations. All authors reviewed the manuscript.

\section*{Additional information}

\textbf{Competing interests} The authors declare no competing interests.


\begin{table}[h]
    \centering
    \caption{\textbf{Writing parameters for serpentine superlattice devices.}  Devices A and B were written on the same canvas (30 $\mathrm{\mu m}$ $\times$ 30 $\mathrm{\mu m}$ area).}
    \begin{tabular}{cccc}
    \hline
    Device & Amplitude (nm) & Wavelength (nm) & Periods \\ \hline
    A      & 5              & 10              & 40      \\
    B      & 5              & 11.5            & 34      \\ \hline
    \end{tabular}
    \label{tab:devices}
\end{table}

\beginsupplement

\end{document}